\documentclass[11pt,a4paper]{article}

\usepackage{epsfig}

\newcommand{\be}{\begin{equation}}
\newcommand{\ee}{\end{equation}}
\newcommand{\bea}{\begin{eqnarray}}
\newcommand{\eea}{\end{eqnarray}}

\begin{document}
\begin{titlepage}

\begin{flushright}
WUE-ITP-2004-007\\
DESY 04--030\\
NSF-KITP-04-29\\
February 2004
\end{flushright}
\vspace{1.cm}

\begin{center}
\large\bf
{\LARGE\bf Numerical evaluation of phase space integrals by sector decomposition}\\[1cm]
\rm
{ T.~Binoth$^{a}$ and  G.~Heinrich$^{b}$}\\[1cm]

{\em $^{a}$Institut f\"ur Theoretische Physik und Astrophysik\\
           Universit\"at  W\"urzburg, Am Hubland, 
	   97074 W\"urzburg, Germany} \\[.5cm]

{\em $^{b}$II. Institut f\"ur theoretische Physik, 
Universit\"at Hamburg,\\ 
Luruper Chaussee 149, 
22761 Hamburg, 
Germany}\\[.5cm]

\end{center}
\normalsize

\vspace{1cm}

\begin{abstract}
In a series of papers we have developed the method of iterated sector 
decomposition for the calculation of infrared divergent
multi-loop integrals. Here we apply it to phase space integrals
 to calculate  a contribution to the double real emission part 
of the $e^+e^-\to 2$\, jets cross section at NNLO. 
The explicit cancellation of infrared poles upon summation
over all possible cuts of a given topology is worked out
in detail for a specific example.
\end{abstract}



\end{titlepage}

\section{Introduction}

The precision measurements at LEP, SLAC, HERA and the Tevatron 
in the past years made it obvious that QCD corrections 
at next-to-leading order (NLO) accuracy are mandatory for a 
successful comparison of data and theory. 
Fortunately, a multitude of NLO corrections are available 
these days and the techniques for these calculations have 
reached a rather mature state, at least in what concerns 
$1\to 3$ or $2\to 2$ partonic processes. 

However, the advent of the LHC and above all an $e^+e^-$ linear collider 
will provide instances where the NLO QCD 
corrections are -- in some particular cases --  
not sufficiently accurate to match the experimental 
precision. This is true for example for the extraction of the 
strong coupling constant $\alpha_s$ from jet observables
at a future $e^+e^-$ collider. 

This fact served as a motivation for substantial progress towards
the calculation of NNLO corrections to important processes, 
in particular in what concerns the two-loop virtual 
corrections\,\cite{m1}--\cite{Glover:2004si} and the one-loop 
corrections combined with 
real radiation where one additional parton is 
emitted\,\cite{onel,kos1,wz1}.
The construction of a fully differential NNLO partonic Monte Carlo
program however also requires the calculation of the real emission part 
where up to two partons can be unresolved, and -- last but not least --
the combination of all the ingredients to a stable and 
sufficiently fast Monte Carlo program. 
In what concerns the double real emission, 
subtraction schemes have been proposed in the 
literature \cite{twot,Kosower:2002su,wz2}, but a complete calculation 
including the final Monte Carlo program  has been performed 
so far only for the particular case of the photon\,+\,jet--
rate in $e^+e^-$ annihilation~\cite{ggam1,ggam}.
However, a lot of activity concerning this subject is going on at the moment.
The efforts are concentrated particularly on the process 
$e^+e^-\to 3$\,jets,  as this reaction 
is both appealing from a phenomenological point of view 
(e.g. measurement of  $\alpha_s$) as well as from a theoretical one 
(no problems due to initial state singularities).  
Nevertheless, it is worthwhile to consider first the reaction 
$e^+e^-\to 2$\,jets as this is the simplest example 
where the treatment of the double unresolved real radiation can be 
studied, and therefore a good testing ground for a new method. 

Progress in what concerns  the integration of subtraction terms
has been made in \cite{Gehrmann-DeRidder:2003bm}, 
where it has been shown that the integrals 
of any $1\to 4$ matrix element in massless QCD 
over the total phase space can be expressed by four master integrals. 
These integrals have been evaluated analytically as well as numerically.  

The  complexity of the phase space integrals with 
two unresolved partons stems from the fact that
the corresponding IR singularity structure is overlapping.  
We have demonstrated in  \cite{Binoth:2000ps} how one can disentangle
in an automated way overlapping singularities
in parameter representations of dimensionally regulated 
multi-loop integrals. The method of {\em iterated sector decomposition}, 
combined with numerical integration of the pole coefficients, 
proved successful to deal with very complicated 
Feynman diagrams \cite{Binoth:2000ps,Binoth:2003ak}. 

In this paper we apply the same ideas to parameter representations
of phase space integrals, as a continuation of the work 
done in \cite{Gehrmann-DeRidder:2003bm,Heinrich:2002rc}. 
In the same context the method of  sector decomposition
also has been applied meanwhile in \cite{Anastasiou:2003gr}.
 
In contrast to\,\cite{Gehrmann-DeRidder:2003bm}, 
where this method of numerical integration has been applied to master integrals 
only, we show here that we also can deal with numerators coming 
from gauge couplings and thus finally with complete matrix elements. 
Working through one example in great detail, we
demonstrate the viability of our approach, showing explicitly the
cancellation of infrared poles for a given topology
when summing over all  cuts.

The paper is organised as follows. 
In section 2 we discuss in detail the cancellation of infrared divergences
in a concrete example. We work out the cut structure and the counterterms
and represent the considered topology by a number of cut diagrams 
which give rise to phase space integrals. In section
3 we present the analytical results for the 2-- and 3--particle cut diagrams, 
and in section 4 we discuss the method to perform the
phase space integration of the 4-particle cut numerically.
In section 5 we present the numerical result for a more complicated
topology which has up to 
$1/\epsilon^4$ poles. Section 6 closes the paper with a discussion.  

\section{Cancellation of infrared divergences}

Two-jet production in $e^+e^-$ collisions at NNLO in $\alpha_s$  
corresponds diagrammatically
to sums over cuts of three-loop vacuum polarisation graphs. 
After renormalization of ultraviolet subdivergences the 
three-loop diagram has at most an overall ultraviolet divergence with
a finite imaginary part. Because this imaginary part 
is related to the sum over all cuts of the diagram, 
one concludes  that the sum over all 
cuts has to be finite. This is just the KLN\,\cite{kln} cancellation mechanism
for infrared final state singularities. 
In our case, one has 2--, 3-- and 4--particle cuts
which correspond to  phase space integrations over the respective partons.
We require 2 jets in the final state, such that up to two partons can become 
unresolved. 

Consider for example the diagram shown in Fig.\,\ref{topo1}.   
\unitlength=1mm
\begin{figure}
\begin{picture}(120,30)
\put(35,0){\epsfig{file=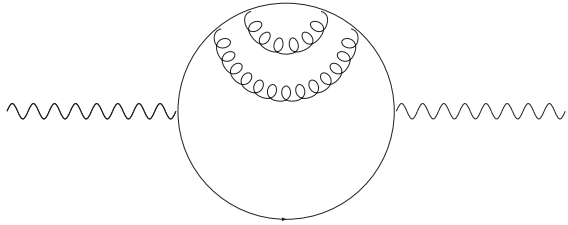,height=3.cm}}
\end{picture}
\caption{\label{topo1}{Sample three loop topology related to 
$e^+e^-\to 2$\,jets at NNLO.}}
\end{figure}
In the following subsections, we will work out the UV 
renormalization and cut structure of this diagram. 
We use Feynman gauge throughout the paper.

\subsection{Cut diagrams}

Infrared cancellations take place when summing over cuts 
of a given renormalized topology. In our case we have
the 2--, 3-- and 4--particle cuts $C_2, C_3$ and $C_4$, 
which graphically are given  by

\unitlength=1mm
\begin{picture}(130,60)
\put(0,39){$C_2 =$ }
\put(51,39){$+\, z_1$}
\put(100,39){$+ \frac{1}{2}\,z_2$}
\put( 10,30){\epsfig{file=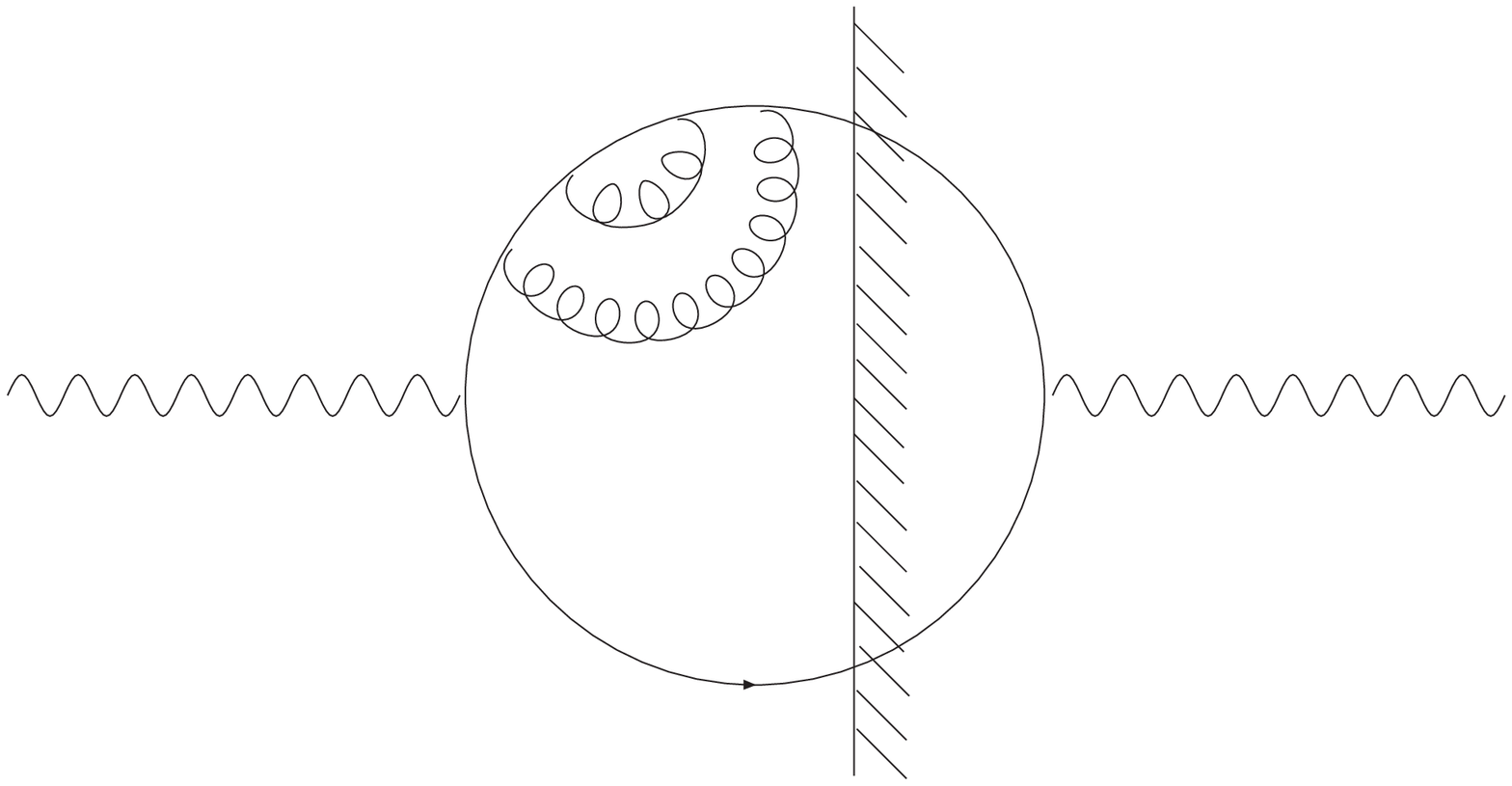,height=2.cm}}
\put( 60,30){\epsfig{file=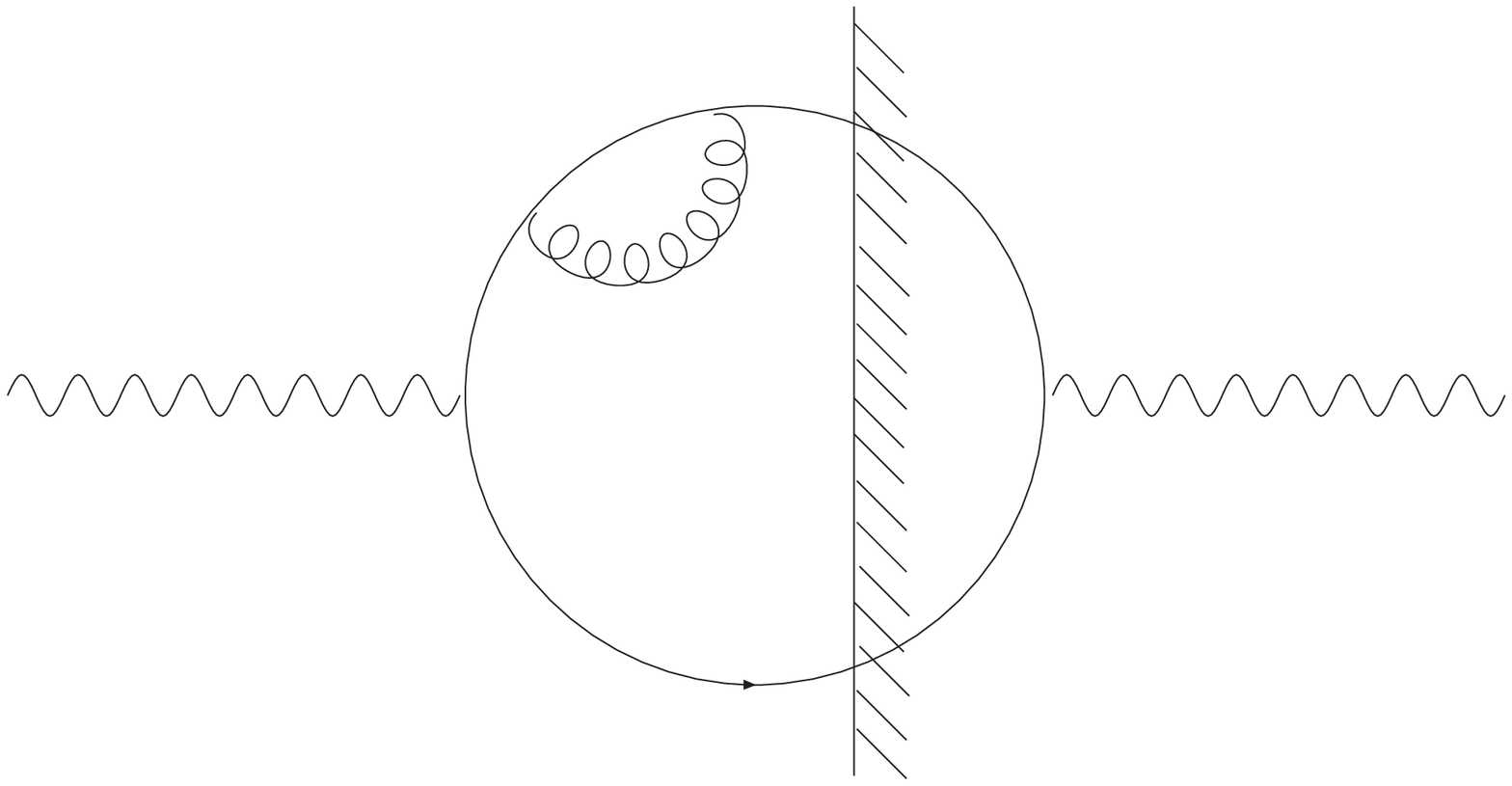,height=2.cm}}
\put(112,30){\epsfig{file=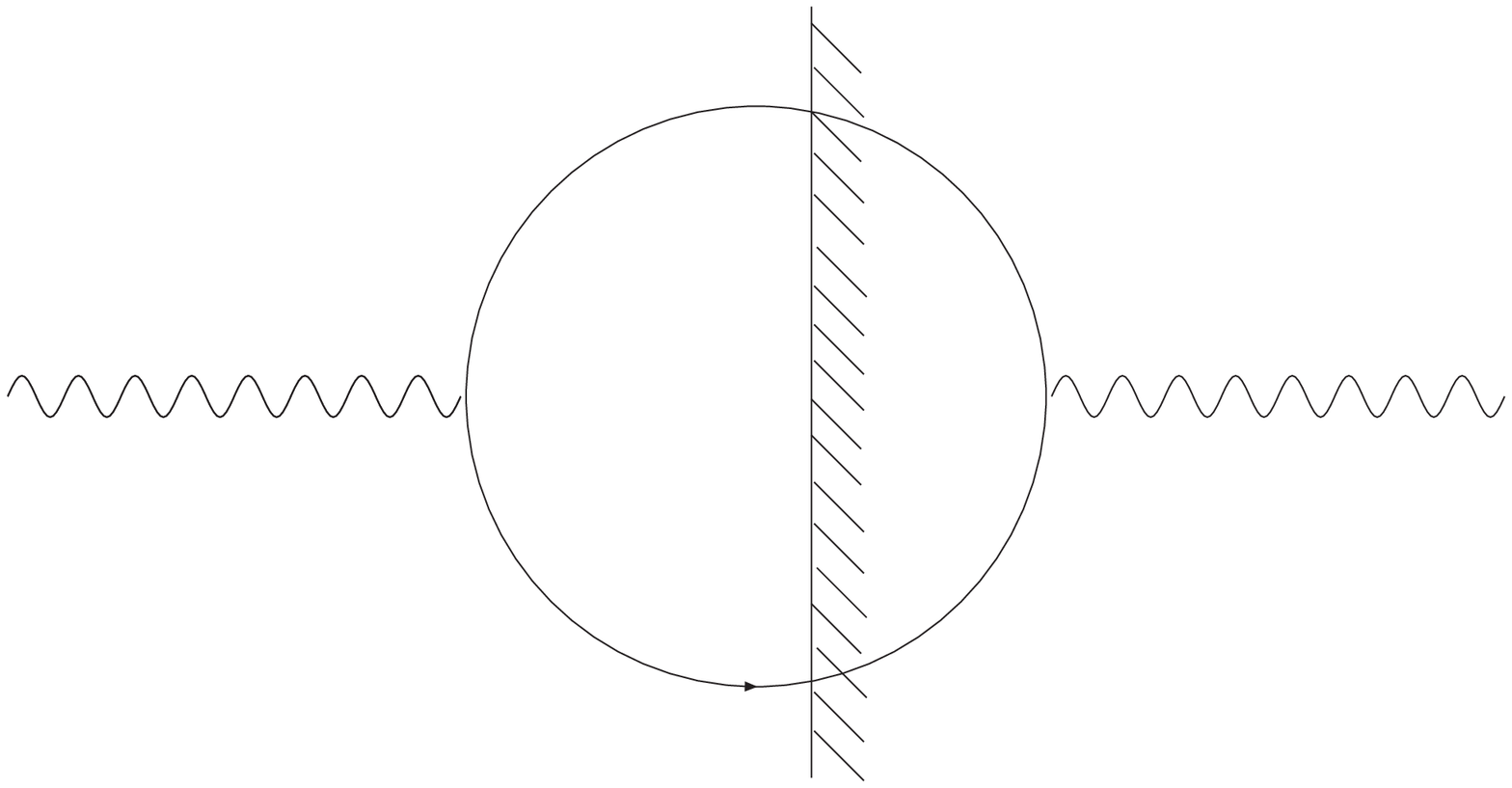,height=2.cm}}
\put(6,9){$+$ }
\put(51,9){$+\, z_1$}
\put(100,9){$+ \frac{1}{2}\, z_2$}
\put( 10,0){\epsfig{file=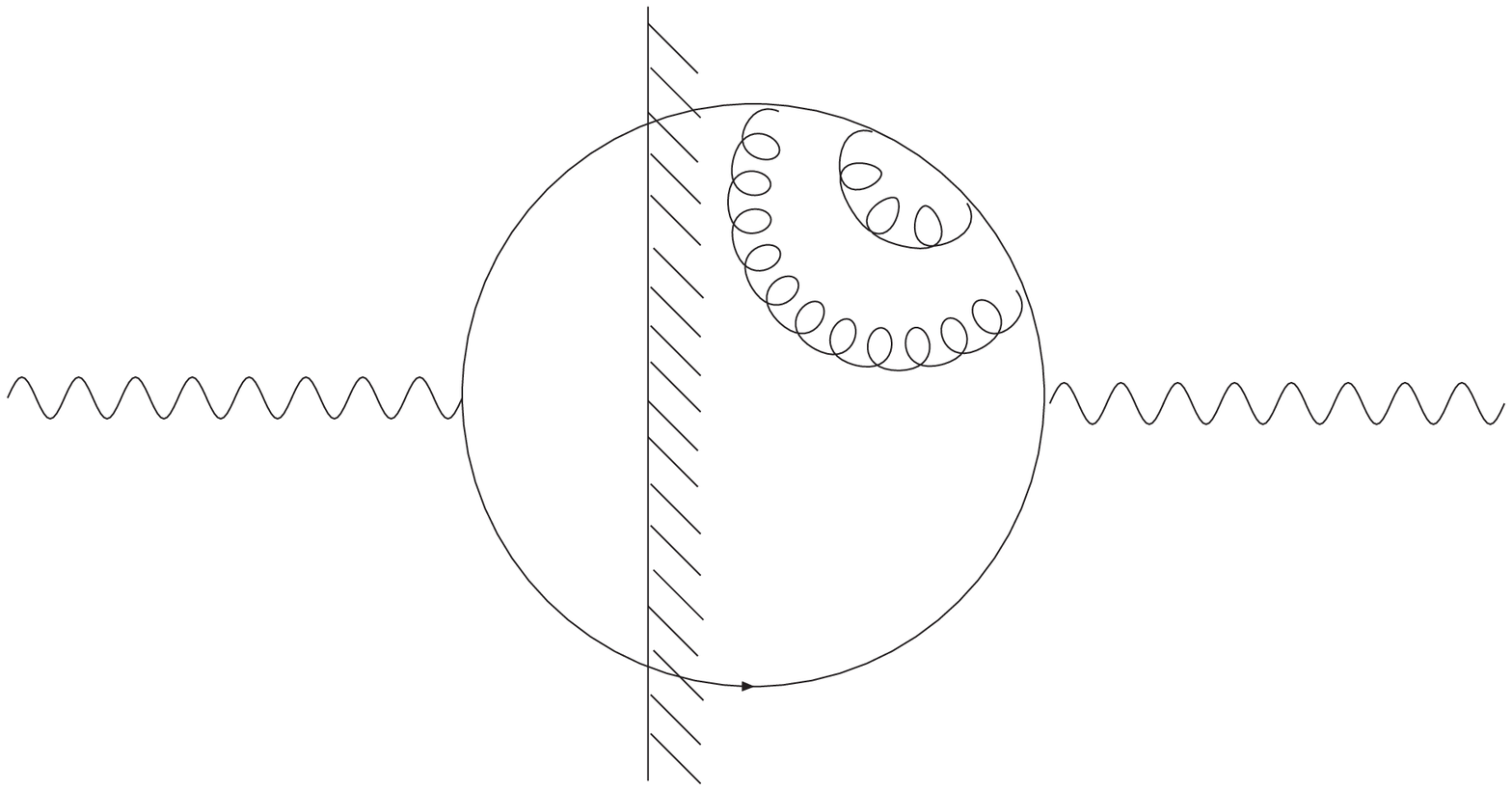,height=2.cm}}
\put( 60,0){\epsfig{file=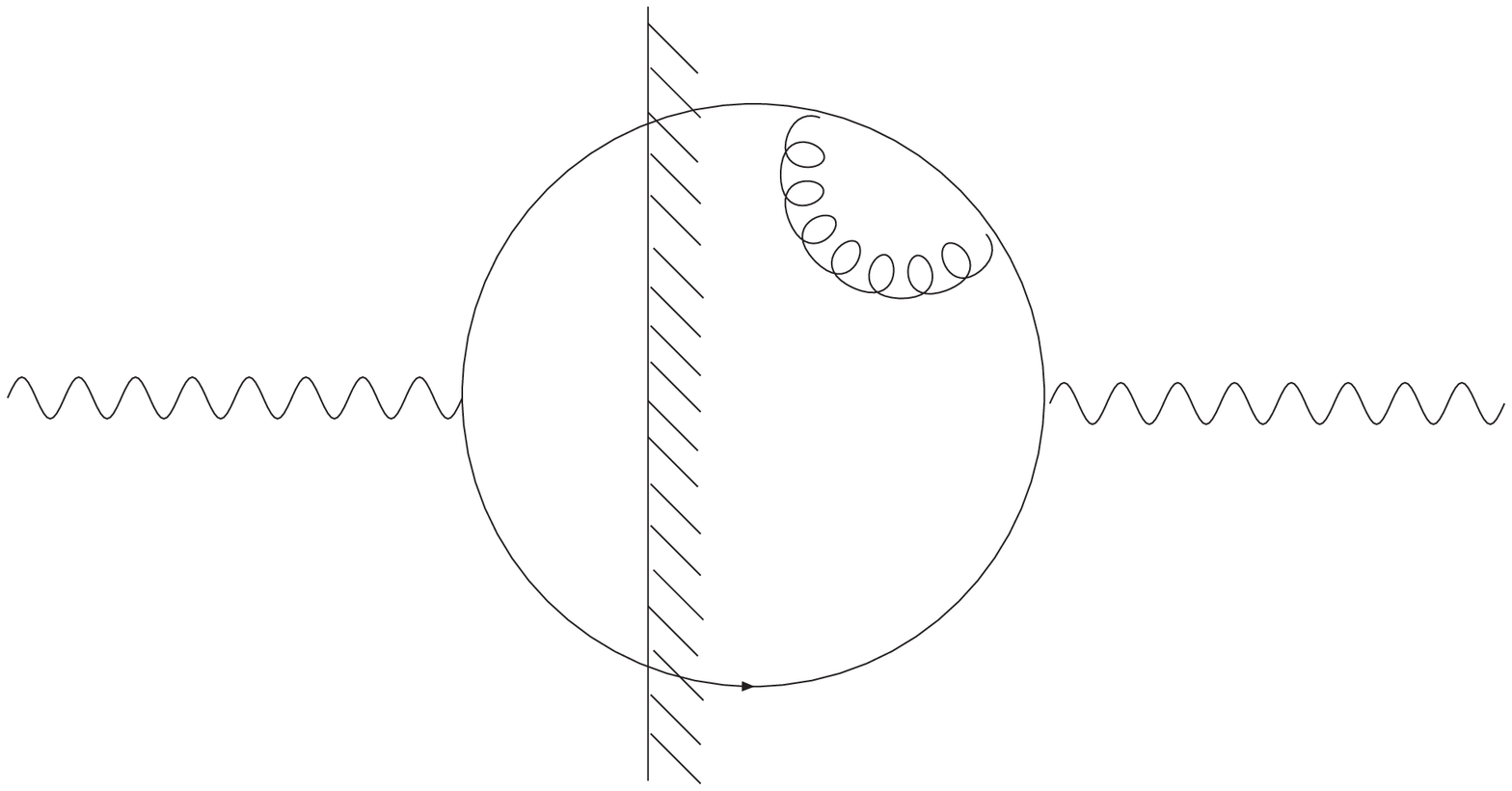,height=2.cm}}
\put(112,0){\epsfig{file=PhaseSpace1to2.eps,height=2.cm}}
\end{picture}

\begin{picture}(120,60)
\put(0,39){$C_3 =$ }
\put(51,39){$+ \frac{1}{2}\,z_1$}
\put( 10,30){\epsfig{file=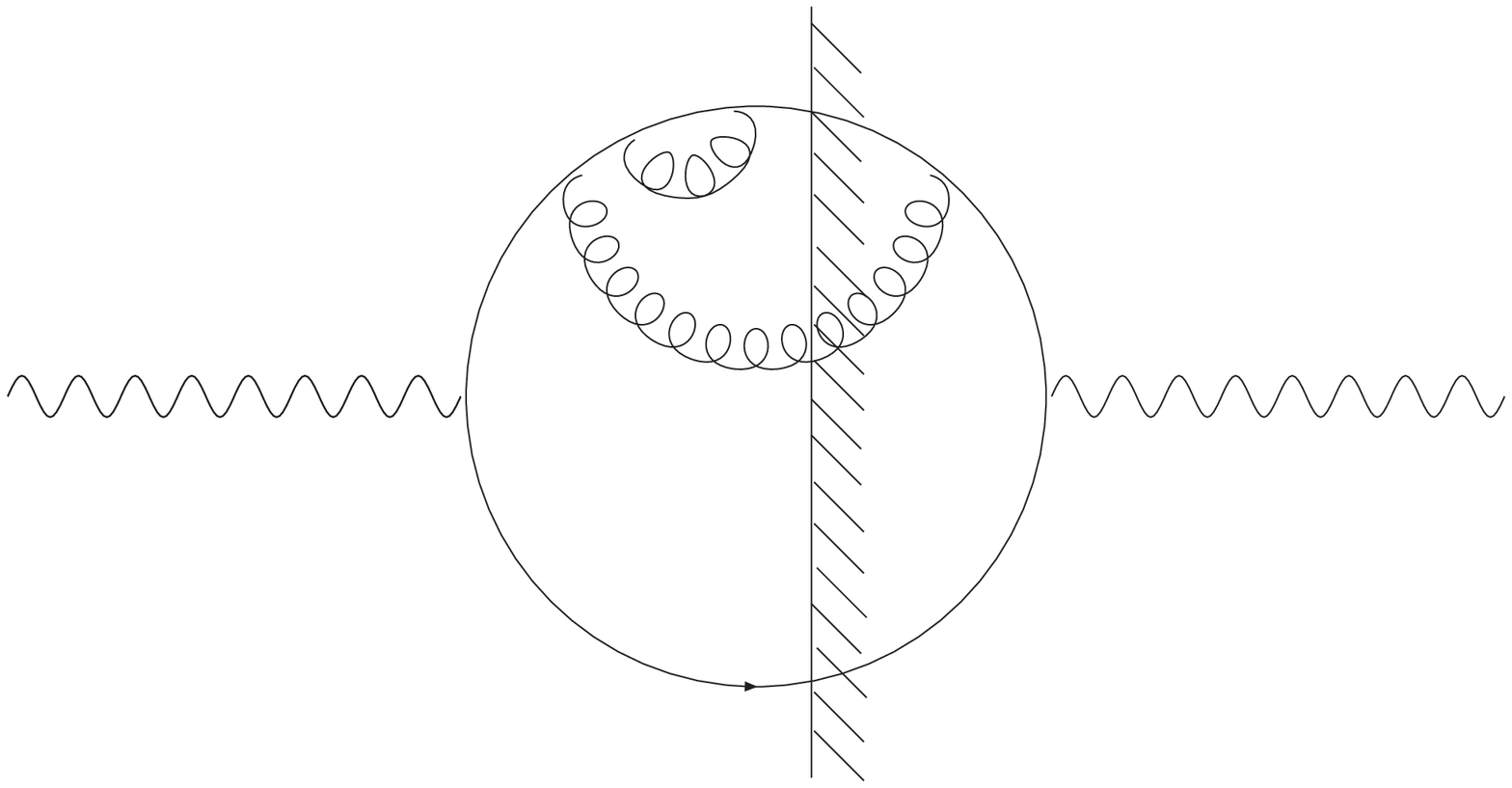,height=2.cm}}
\put( 62,30){\epsfig{file=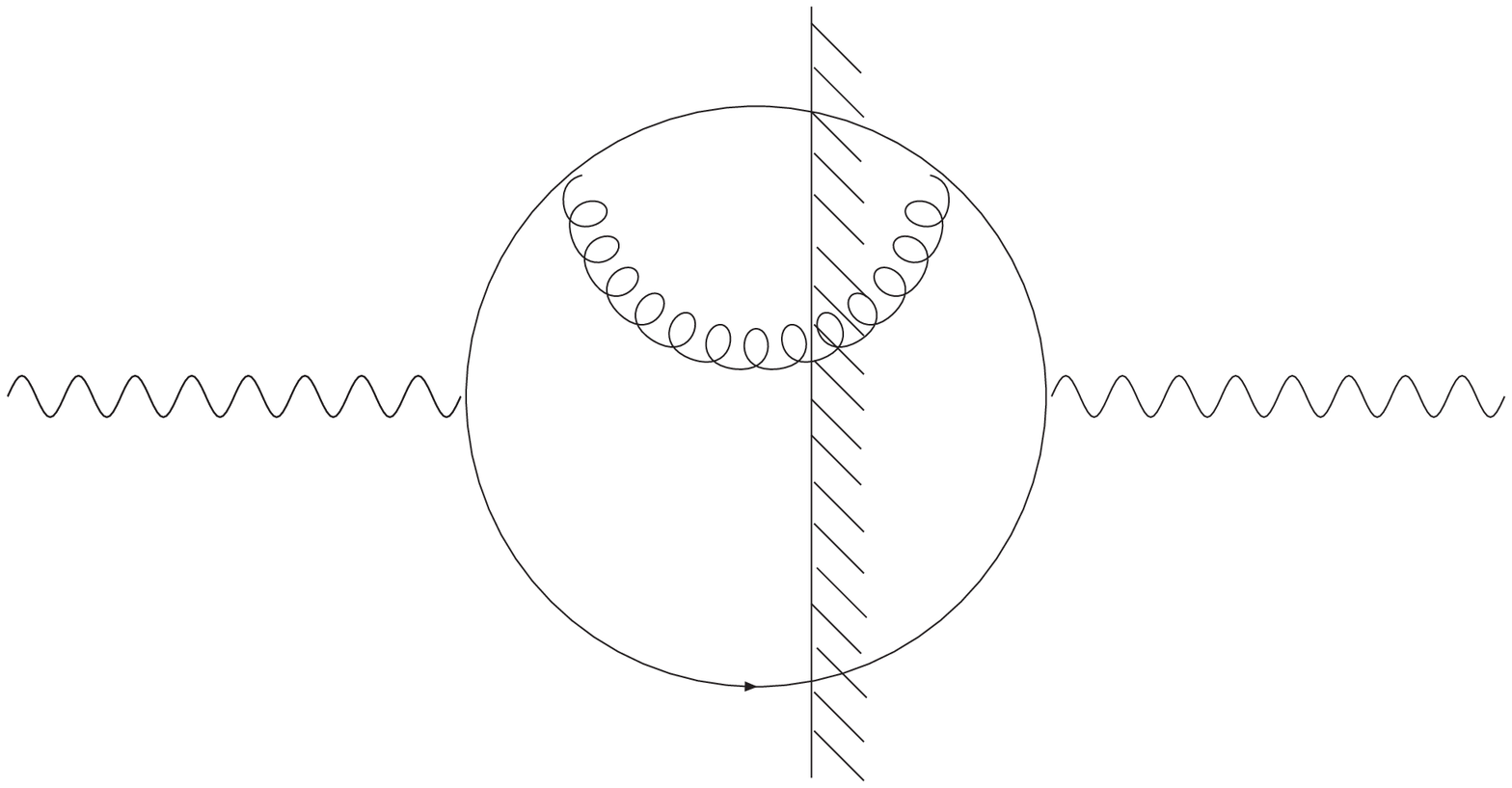,height=2.cm}}
\put(6,9){$+$ }
\put(51,9){$+ \frac{1}{2}\, z_1$}
\put( 10,0){\epsfig{file=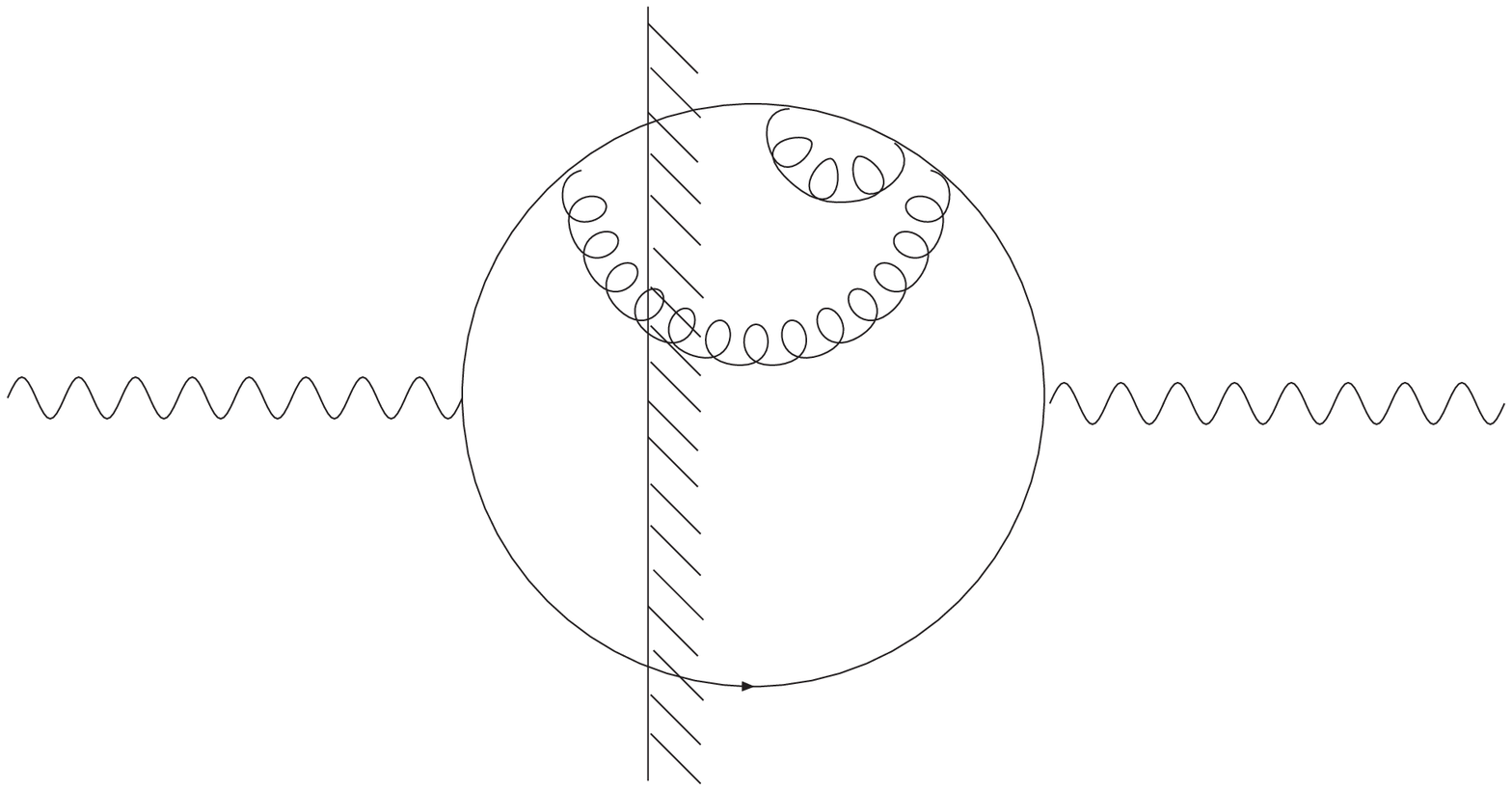,height=2.cm}}
\put( 62,0){\epsfig{file=PhaseSpace1to3.eps,height=2.cm}}
\end{picture}

\begin{picture}(120,30)
\put(0,9){$C_4 =$ }
\put(10,0){\epsfig{file=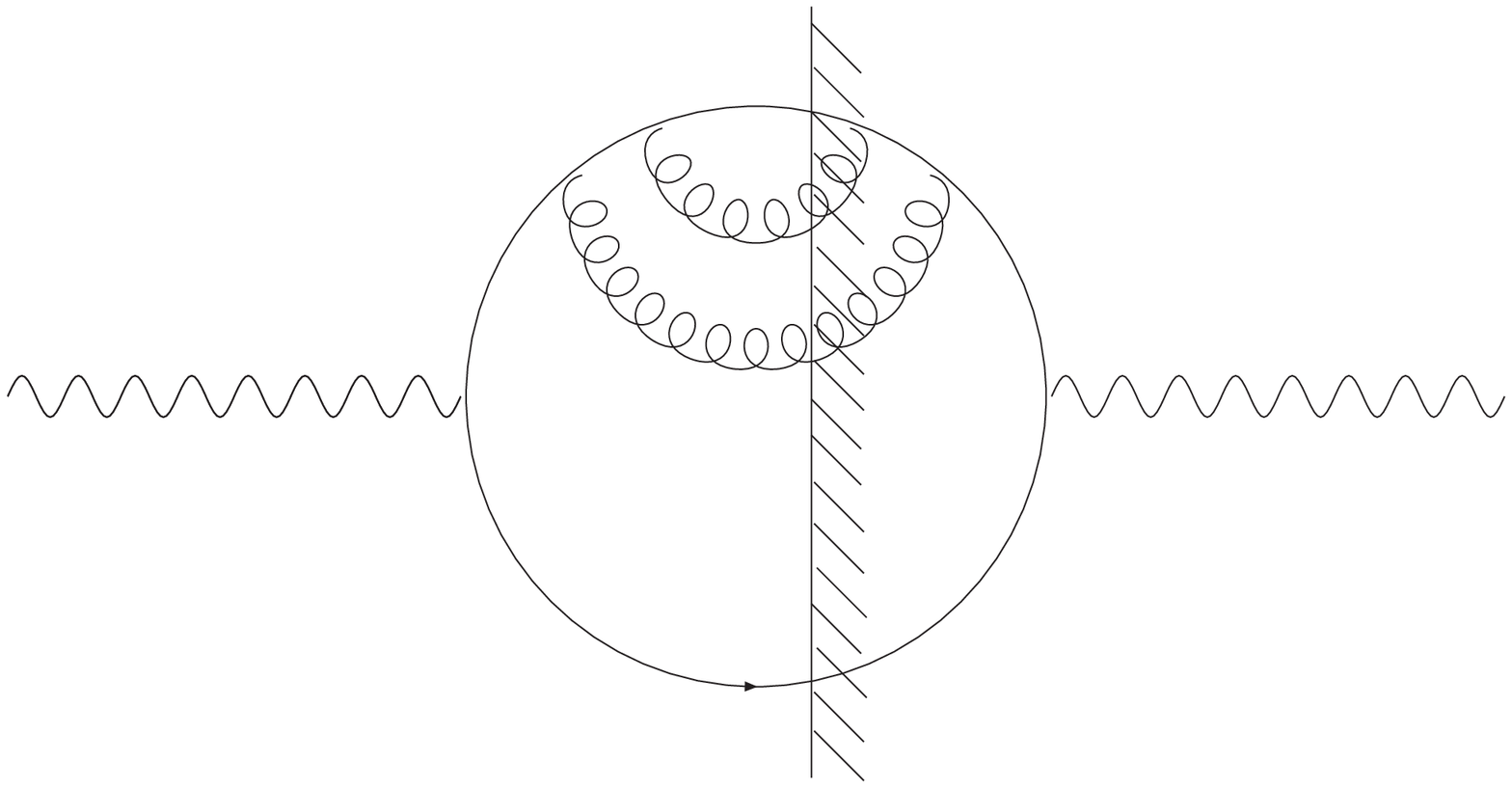,height=2.cm}}
\end{picture}

The  factors $z_1$ and $z_2$ denote the contributions from counterterms.
The sum of terms in each line in the above figures
is formally UV finite. 
In dimensional regularization 
the radiative corrections to the two-point functions vanish if the 
particle is put on-shell. This is the mechanism which formally leads
to the conversion of UV into IR poles.
Using the diagrammatic rule 

\unitlength=1mm
\begin{picture}(120,15)
\put(50,3){$= 0\,\,,$}
\put(70,3){$= 0$}
\put(10,0){\epsfig{file=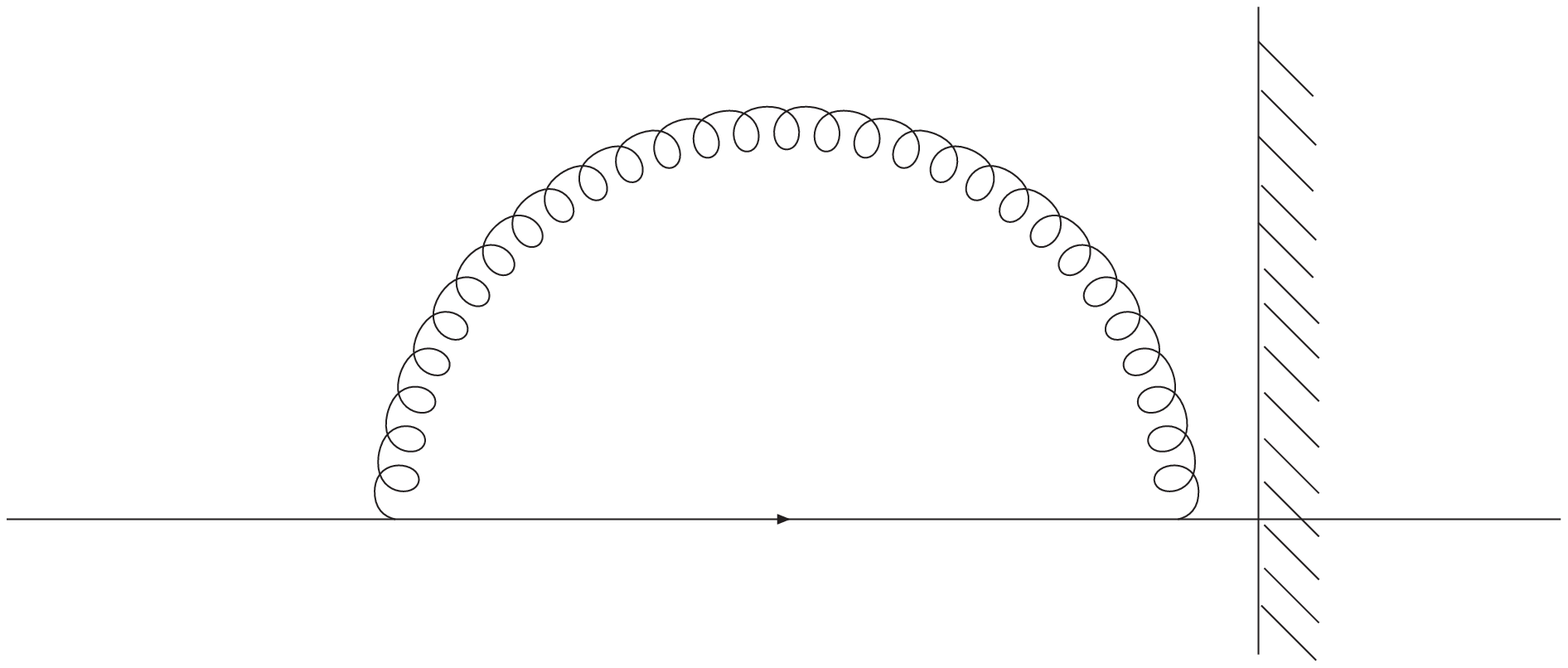,height=1.5cm}}
\put(70,0){\epsfig{file=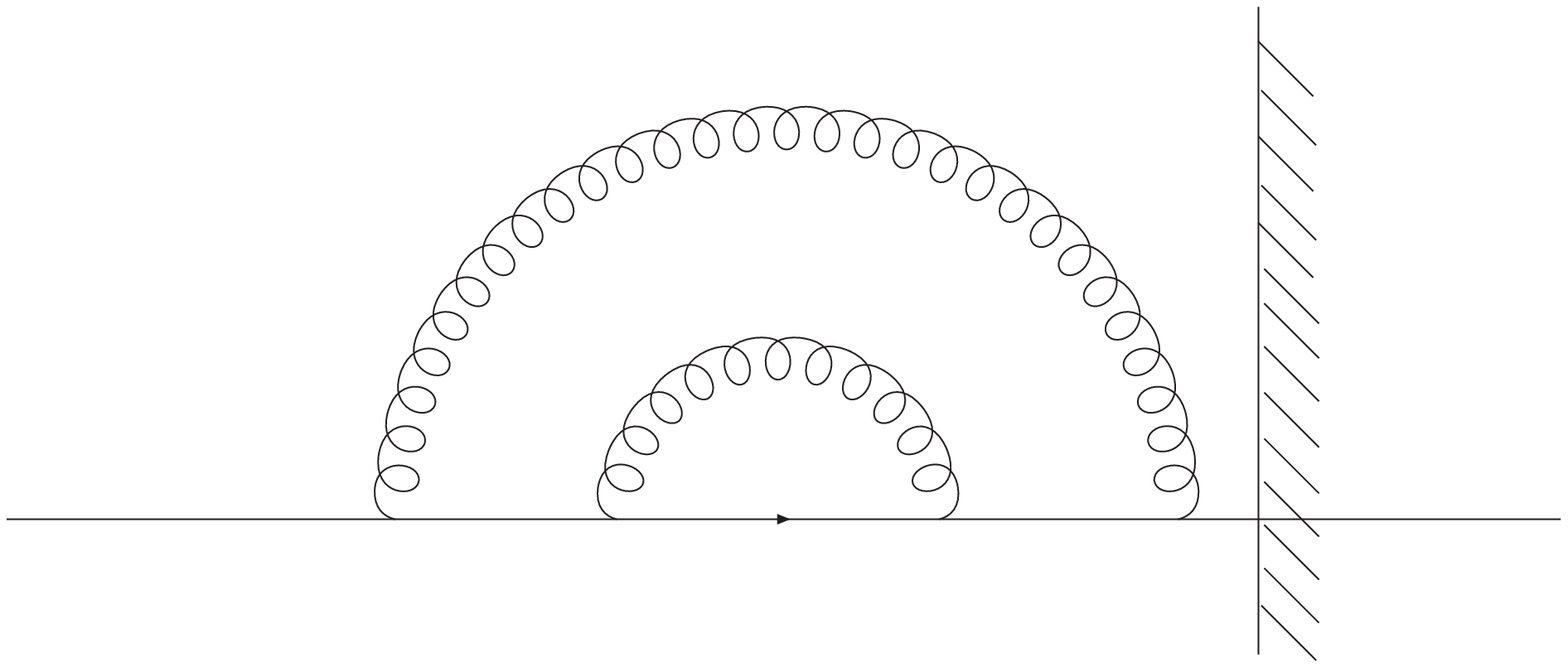,height=1.5cm}}
\put(120,4){$= 0$}
\end{picture}

the only remaining terms in the sum $C=C_2+C_3+C_4$ are 

\unitlength=1mm
\begin{picture}(140,30)
\put(0,9){$C =$ }
\put(51,9){$+\, z_1$}
\put(100,9){$+\, z_2$}
\put( 10,0){\epsfig{file=PhaseSpace1to4.eps,height=2.cm}}
\put( 60,0){\epsfig{file=PhaseSpace1to3.eps,height=2.cm}}
\put(110,0){\epsfig{file=PhaseSpace1to2.eps,height=2.cm}}
\end{picture}
\be\label{EqC}
\ee

The $1/\epsilon$ terms contained in $z_1$ and $z_2$ now represent IR poles. 
The graphs shown in (\ref{EqC})
denote the contribution from the given topology to the full process. 
As the imaginary part of the corresponding 3-loop 
topology is finite, unitarity implies that $C$ is IR finite. 

The part which is  hard to calculate analytically is the 4--particle cut, 
that is why we advocate a numerical evaluation of $C_4$. 
The remaining combination of 
counterterms and 2-- and 3--particle phase space integrations can 
straightforwardly  be done analytically. 
This will be worked out in the next section. 
 
\subsection{UV renormalization}
To compute the renormalization constants 
$z_1$ and $z_2$ we have to determine the
pole parts of all graphs and subgraphs. 
Following a graphical BPHZ notation one has
to carry out the following subtractions to renormalize the 
one-loop selfenergy and finally the whole diagram: 
The one-loop subtraction:

\unitlength=1mm
\begin{picture}(120,20)
\put(61,5){$-$}
\put(10,6){\epsfig{file=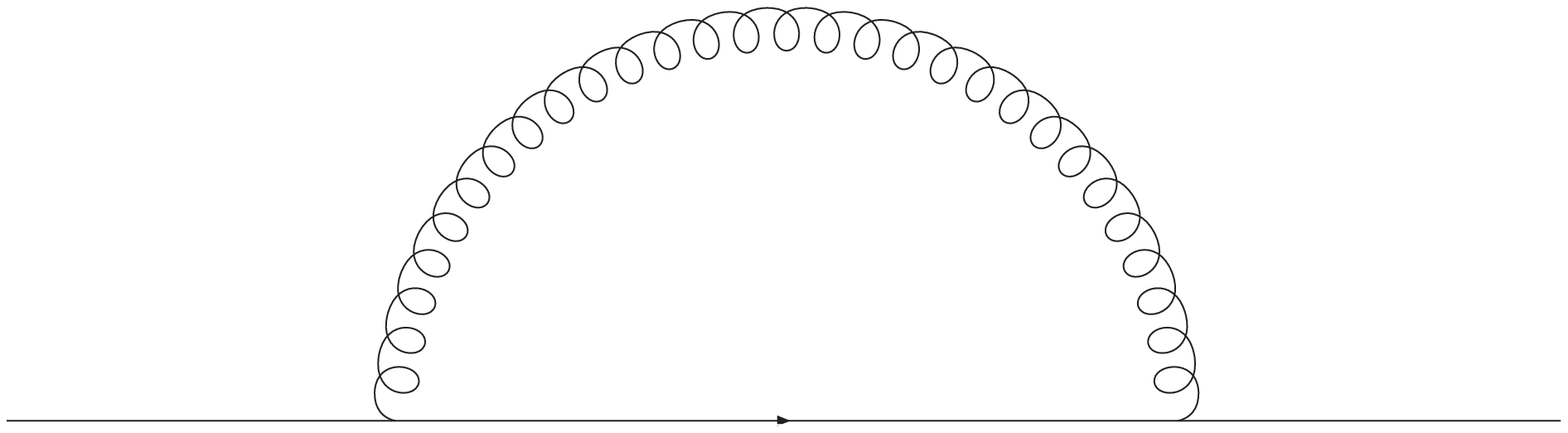,height=1.3cm}}
\put(70,0){\epsfig{file=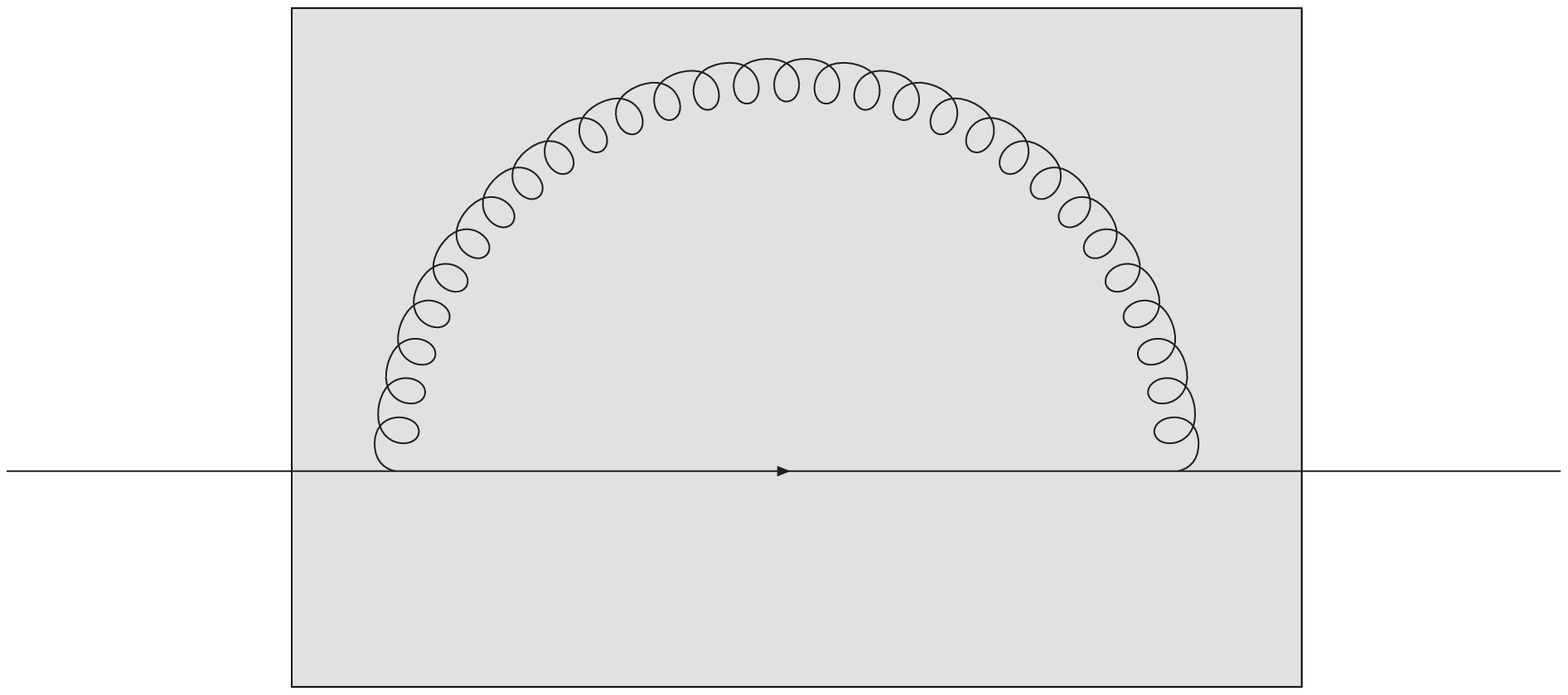,height=2.cm}}
\put(120,5){$=$ finite}
\end{picture}

and the two-loop subtractions:

\unitlength=1mm
\begin{picture}(120,50)
\put(1,6){$-$}
\put(62,6){$+$}
\put(62,27){$-$}
\put(10,28){\epsfig{file=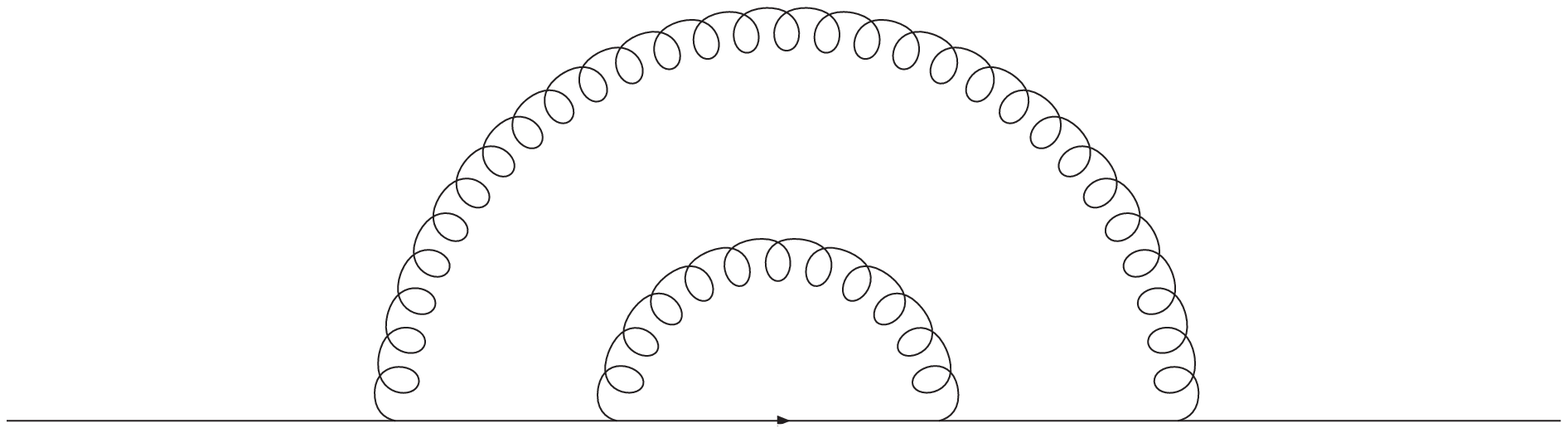,height=1.3cm}}
\put(70,25){\epsfig{file=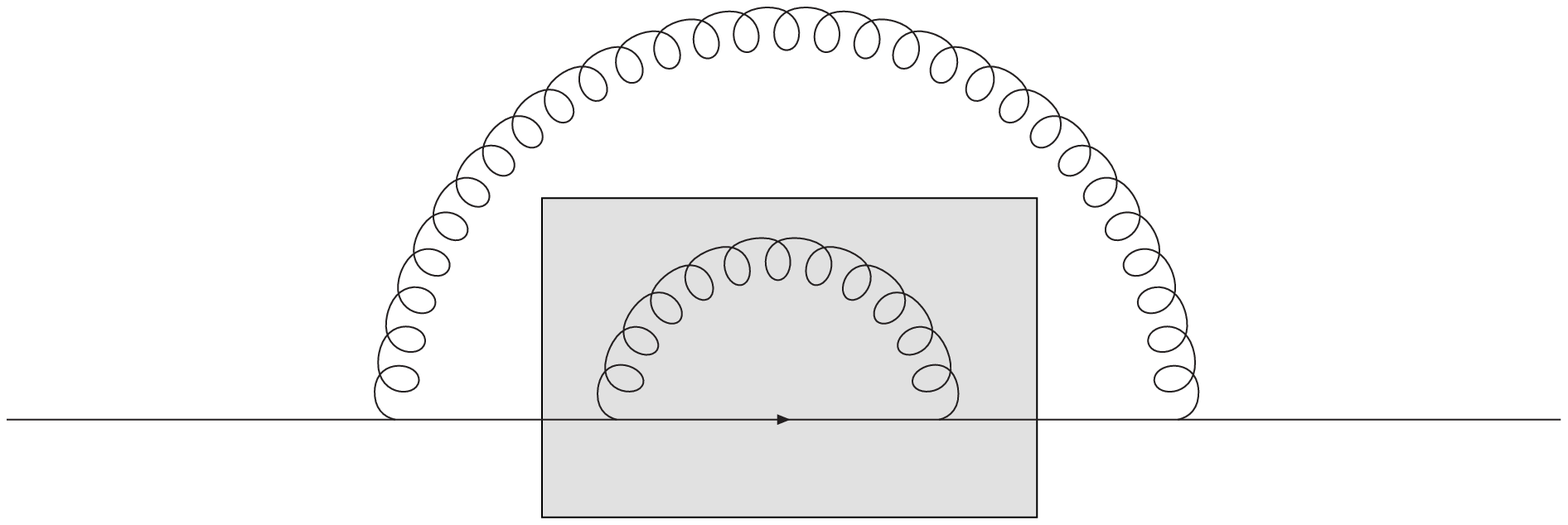,height=1.5cm}}
\put(10,0){\epsfig{file=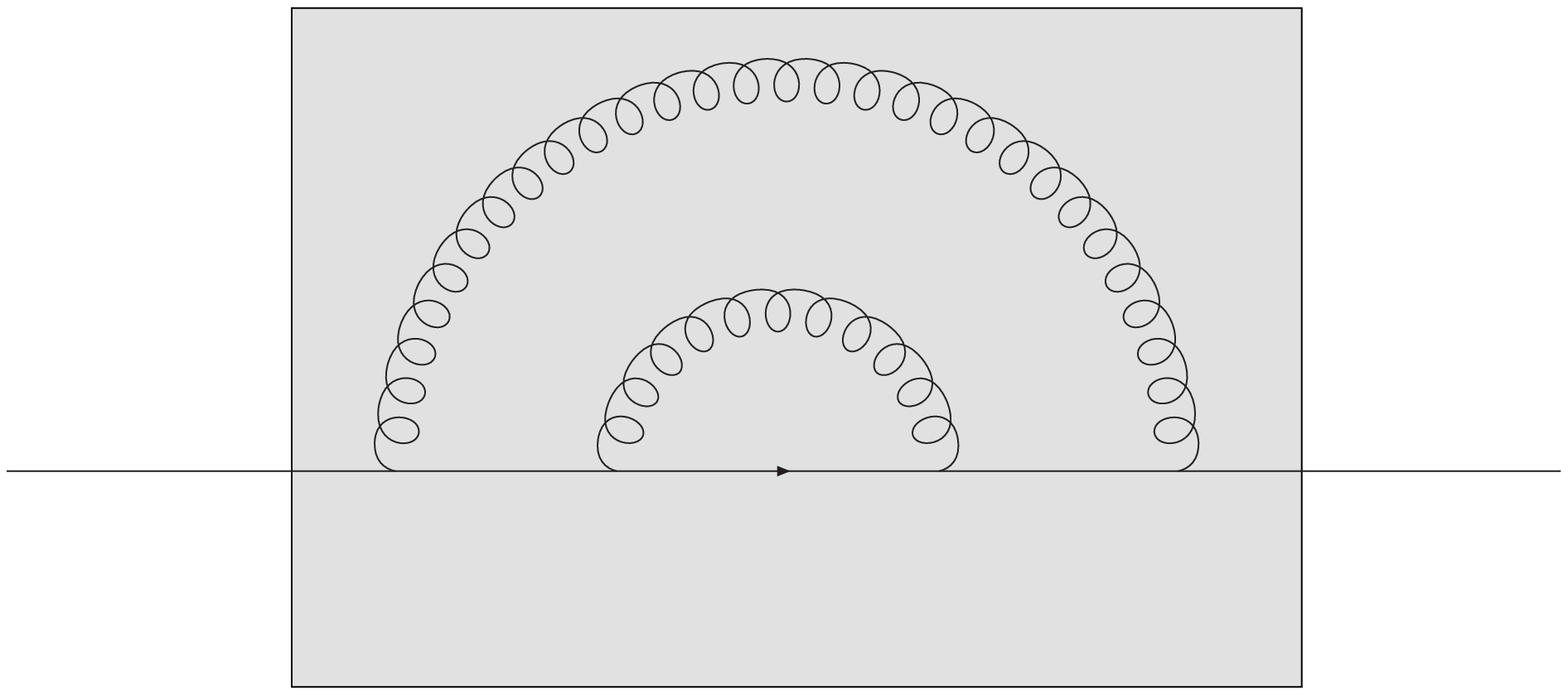,height=2.cm}}
\put(70,0){\epsfig{file=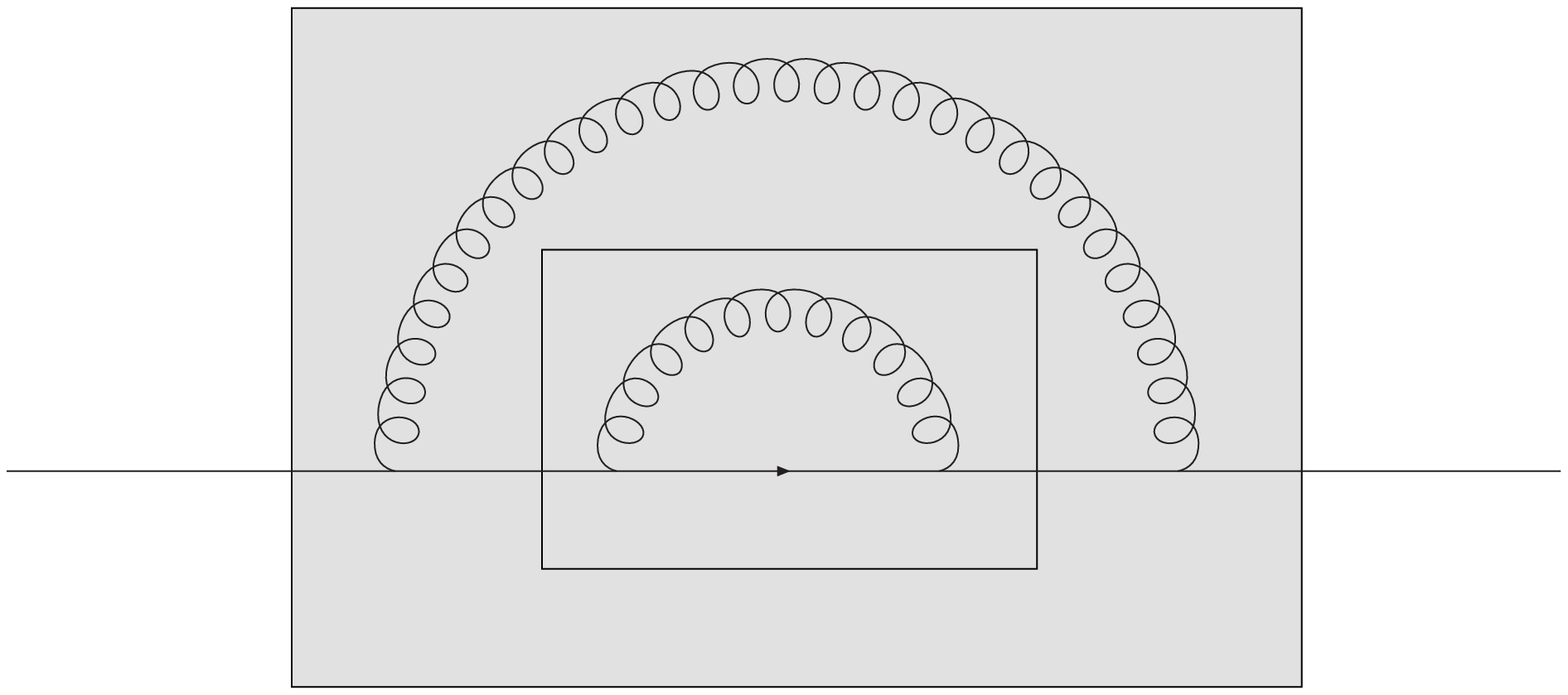,height=2.cm}}
\put(120,4){$=$ finite}
\end{picture}

The shaded boxes denote the $\overline{\rm{MS}}$ prescription to
keep only the pole part of the expression,
up to absorption of a standard factor into the coupling:
\be
\alpha = C_{\overline{\rm{MS}}}\,\alpha_0\, ,\quad 
\alpha_s = C_{\overline{\rm{MS}}}\,\alpha_{s,0} \, ,\quad
C_{\overline{\rm{MS}}}=\frac{ \Gamma(1+\epsilon) (4\pi)^\epsilon}{\mu^{2\epsilon}}
\ee
Using the integral
\bea
&&I(\alpha,\beta) = \int \frac{d^Dk}{i\pi^{D/2}} 
\frac{\gamma^\mu \slash \!\!\!k\gamma_\mu}{[-k^2]^{\alpha}[-(k-p)^2]^{\beta}} 
\nonumber\\
&&= \slash \!\!\!p (2-D)(-p^2)^{D/2-\alpha-\beta}
\frac{\Gamma(\alpha+\beta-D/2)}{\Gamma(\alpha)\Gamma(\beta)}
\frac{\Gamma(D/2-\alpha+1)\Gamma(D/2-\beta)}{\Gamma(D-\alpha-\beta+1)}
\eea
one finds the following analytic results for the graphical 
expressions ($D=4-2\epsilon$).
At one loop:
\bea
\includegraphics[width=2.5cm]{FermionSelfEnergy_1l.eps} 
&=& i \slash \!\!\!p \,C_F
\left( \frac{\alpha_s}{4\pi} \right)
\left( \frac{-p^2}{\mu^2} \right)^{-\epsilon}
\frac{\Gamma(1-\epsilon)^2}{\Gamma(2-2\epsilon)}
\frac{(1-\epsilon)}{\epsilon}\nonumber\\
\includegraphics[width=2.5cm]{FermionSelfEnergy_1l_c.eps} 
&=& i \slash \!\!\!p \,C_F
\left( \frac{\alpha_s}{4\pi} \right) \frac{1}{\epsilon} 
\eea
and for the  two-loop part:
\bea
\includegraphics[width=2.5cm]{FermionSelfEnergy_2l.eps} 
&=& -i \slash \!\!\!p \,C_F^2
\left( \frac{\alpha_s}{4\pi} \right)^2
\left( \frac{-p^2}{\mu^2} \right)^{-2\epsilon}
\frac{\Gamma(1+2\epsilon)}{\Gamma(1+\epsilon)^2}
\frac{\Gamma(1-\epsilon)^3}{\Gamma(3-3\epsilon)}
\frac{(1-\epsilon)^2}{\epsilon^2}\nonumber\\
\includegraphics[width=2.5cm]{FermionSelfEnergy_2l_c1.eps} 
&=& -i \slash \!\!\!p \,C_F^2
\left( \frac{\alpha_s}{4\pi} \right)^2
\left( \frac{-p^2}{\mu^2} \right)^{-\epsilon}
\frac{\Gamma(1-\epsilon)^2}{\Gamma(1-2\epsilon)}
\frac{1-\epsilon}{\epsilon^2(1-2\epsilon)}\nonumber\\
\includegraphics[width=2.5cm]{FermionSelfEnergy_2l_c2.eps} 
&=& -i \slash \!\!\!p \,C_F^2
\left( \frac{\alpha_s}{4\pi} \right)^2
\Bigl[ 
       \frac{1}{2\epsilon^2} 
     + \frac{5}{4\epsilon} 
     - \frac{1}{\epsilon}\log\left(\frac{-p^2}{\mu^2} \right)\Bigr]\nonumber\\
\includegraphics[width=2.5cm]{FermionSelfEnergy_2l_c3.eps} 
&=& -i \slash \!\!\!p \,C_F^2
\left( \frac{\alpha_s}{4\pi} \right)^2
\Bigl[ 
       \frac{1}{\epsilon^2} 
     + \frac{1}{\epsilon} 
     - \frac{1}{\epsilon}\log\left(\frac{-p^2}{\mu^2} \right)\Bigr]
\eea
The renormalization constants can be read off directly:
\bea
i \slash \!\!\!p z_1 &=& 
 \includegraphics[width=2.5cm]{FermionSelfEnergy_1l_c.eps}
 \Rightarrow z_1 =  C_F \frac{\alpha_s}{4\pi}  \frac{1}{\epsilon}          \\
i \slash \!\!\!p z_2 &=&  \includegraphics[width=2.5cm]{FermionSelfEnergy_2l_c2.eps}
                          - \includegraphics[width=2.5cm]{FermionSelfEnergy_2l_c3.eps}\nonumber\\ 
 \Rightarrow z_2 &=& C_F^2 \left( \frac{\alpha_s}{4\pi} \frac{1}{\epsilon}\right)^2 
  \Bigl[ 
   \frac{1}{2} -\frac{1}{4}\epsilon
  \Bigr] = \frac{1}{2}\;  z_1^2 \; \Bigl[  1 - \frac{1}{2}\epsilon \Bigr]
\eea
The non-local logarithmic terms cancel, as guaranteed by the BPHZ theorem.

\section{Matrix elements and phase space integrals}
What remains to be done is the evaluation of the 
phase space integrals for the cuts shown in Eq.~(\ref{EqC}).
The corresponding matrix elements are given by
the following formulae, where  
$p_1$ ($p_2$) are the momenta of the quark (anti-quark), 
$p_3$  and $p_4$ denote the gluons.
\bea\label{MEs}
| {\cal M}_{1\to 2} |^2&=& 16\pi\, \alpha\, (1-\epsilon)\,
\left(\frac{\mu^{2\epsilon}}{\Gamma(1+\epsilon)(4\pi)^\epsilon}\right)\,s_{12}\\
| {\cal M}_{1\to 3} |^2&=& 8 \,(4\pi)^2 (1-\epsilon)^2\, \alpha\, \alpha_{s}\, C_F 
\left(\frac{\mu^{2\epsilon}}{\Gamma(1+\epsilon)(4\pi)^\epsilon}\right)^2\;\frac{s_{23}}{s_{13}}\\
| {\cal M}_{1\to 4} |^2&=& 16\, (4\pi)^3 (1-\epsilon)^3 \alpha \,\alpha_{s}^2 \,C_F^2
\left(\frac{\mu^{2\epsilon}}{\Gamma(1+\epsilon)(4\pi)^\epsilon}\right)^3 \nonumber\\&&
\frac{(s_{13}+s_{34})(s_{12}+s_{24}) -
s_{14}\,s_{23}}{s_{13}(s_{13}+s_{34}+s_{14})^2}\label{10}
\eea 
$| {\cal M}_{1\to 2} |^2$ and $| {\cal M}_{1\to 3} |^2$ can be integrated 
directly. The corresponding phase space integrals are 
given in detail in the Appendix. We obtain
\bea
T_{1\to 2} &=& \int d\Phi_{1\to2} | {\cal M}_{1\to 2} |^2 
           = 2 \,\alpha\, Q^2\, \left( \frac{Q^2}{\mu^2}
	   \right)^{-\epsilon} \,
	  \frac{(1-\epsilon)\Gamma(1-\epsilon)}{\Gamma(1+\epsilon)\Gamma(2-2\epsilon)} 
\eea
Using the integral
\bea
J_3(\epsilon) &=& \int\limits_0^{\infty} dy_1 dy_2dy_3\, \delta(1-y_1-y_2-y_3)\, (y_1\,y_2\,y_3)^{-\epsilon}
\frac{y_3}{y_2} =  -\frac{(1-\epsilon)}{\epsilon}
\frac{\Gamma(1-\epsilon)^3}{\Gamma(3-3\epsilon)}\;,
\eea
where $y_1=s_{12}/Q^2, y_2=s_{13}/Q^2,y_3=s_{23}/Q^2$, one finds for the $1\to 3$ case:
\bea
T_{1\to 3} &=& \int d\Phi_{1\to3} | {\cal M}_{1\to 3} |^2\nonumber\\
&=& \alpha\, \alpha_s\, C_F \,\frac{(2 \pi)^{2\epsilon}}{8\pi^3}
(1-\epsilon)^2 Q^2 \left( \frac{Q^2}{\mu^2} \right)^{-2\epsilon}
\frac{V(3-2\epsilon)\,V(2-2\epsilon)}{\Gamma(1+\epsilon)^2} \, J_3(\epsilon)\nonumber\\
&=& -z_1\, T_{1\to 2}\,\left( \frac{Q^2}{\mu^2} \right)^{-\epsilon} 
 \frac{2\,(1-\epsilon)^2\Gamma(1-\epsilon)^2}{\Gamma(1+\epsilon)\Gamma(3-3\epsilon)}
 \eea
The $1\to 4 $ case cannot be solved easily analytically. We write it in the following form:
\bea
T_{1\to 4} &=& \int d\Phi_{1\to4} | {\cal M}_{1\to 4} |^2\nonumber\\
&=&  ( z_1\, \epsilon )^2\, T_{1\to 2}\,\left( \frac{Q^2}{\mu^2} \right)^{-2\epsilon}
 \frac{(1-\epsilon)^2 }{\Gamma(1+\epsilon)^2\Gamma(1-2\epsilon)} \, J_4(\epsilon)
\eea
where the  nontrivial integral $J_4(\epsilon)$ remains, which will be dealt with numerically in the next 
section. 

So far, one finds for the  cut diagrams in Eq.\,(\ref{EqC}):
\bea
C_2 &=& z_1^2\, T_{1\to 2}\,\left( \frac{1}{2} - \frac{1}{4}\epsilon 
\right)\nonumber\\
C_3 
&=&-z_1^2\, T_{1\to 2}\,\left( \frac{Q^2}{\mu^2} \right)^{-\epsilon}
\,(1-\epsilon)^2\left[1+\frac{9}{2}\epsilon+
\epsilon^2\left(\frac{63}{4}-\frac{2\pi^2}{3}  \right)+{\cal O}(\epsilon^3)\right]\nonumber\\
C_4 
&=&z_1^2\, T_{1\to 2}\left( \frac{Q^2}{\mu^2} \right)^{-2\epsilon}  \epsilon^2 (1-\epsilon)^2 
\left[1-\epsilon^2\frac{\pi^2}{2}
+{\cal O}(\epsilon^3)\right]\, J_4(\epsilon) 
\label{EqC4}
\eea
and consequently the poles contained in $J_4(\epsilon)$ have to cancel with
\bea
\Bigl( C_2+C_3 \Bigr)\vert_{\rm pole\, part} = z_1^2\; T_{1\to 2} \; 
 \left[ 
   -\frac{1}{2} - \frac{11}{4} \epsilon + \epsilon\; \log\left( \frac{Q^2}{\mu^2}\right) 
 \right]\;.
 \label{c23}
\eea
\section{Numerical evaluation of $1\to 4$ phase space integrals}

In massless QCD, the integrals of any $1\to 4$ matrix element over 
the total phase space can be expressed by four master integrals 
whose analytical evaluation is complicated, 
but has been achieved in\,\cite{Gehrmann-DeRidder:2003bm}. 
This allows to follow the conventional procedure to establish a subtraction
scheme and to integrate the subtraction terms analytically. 
However, as the finite part of the phase space will finally be 
integrated numerically anyway,  
a flexible, completely numerical method would be welcome. 
Of course the problem consists in the isolation and subtraction of 
the infrared poles, stemming from the integration over unresolved 
particles, before a numerical evaluation is possible. 
This is where sector decomposition is very convenient. 
How it proceeds is shown for the sample integral $J_4$ 
defined in eq.~(\ref{EqJ4}). 
Note that this integral  contains 
an integrable singularity of square-root type. 
For the sake of numerical  stability it is preferable to perform a 
mapping 
such that  the integrand is bounded near the phase space limits.
In detail we proceed as follows:

Our starting point is the integral $J_4(\epsilon)$ which represents (up to 
an overall factor, see eqs.\,(\ref{10}),(\ref{EqC4})) 
the integral over the 4--particle cut of the topology in Fig.~\ref{topo1}.
\bea\label{EqJ4}
J_4(\epsilon) = \frac{4}{\pi}\int\limits_{0}^{\infty} \prod_{i=1}^6d\,y_i\, 
\Theta(-\lambda)  (-\lambda)^{-1/2-\epsilon} \delta(1-\sum_{j=1}^{6} y_j)
 \frac{(y_1+y_5)\,(y_2+y_6)-y_3\,y_4}{y_2 \,(y_2+y_4+y_6)^2}\;,
\eea
where we have rescaled the Mandelstam invariants
by \begin{displaymath}
y_1=s_{12}/Q^2, y_2=s_{13}/Q^2, 
y_3=s_{23}/Q^2,
y_4=s_{14}/Q^2,y_5=s_{24}/Q^2,
y_6=s_{34}/Q^2
\end{displaymath}
and $\lambda\equiv\lambda(y_1y_6,y_2y_5,y_3y_4)$ is the K\"allen function 
$\lambda(x,y,z)=x^2+y^2+z^2-2xy-2yz-2xz$. 
The derivation of the phase space integral can be found in the appendix, 
see eq.\,(\ref{fi4}). 

\subsection*{Primary sector decomposition}

To eliminate the delta distribution in (\ref{EqJ4}) we split the  
integration region into 6 "primary sectors" by the following decomposition 
of unity:
\be
1 = \sum\limits_{j=1}^6 \prod\limits_{j\not =k=1}^{6}\Theta(y_j-y_k)
\ee     
In each primary sector $j$ we apply the mapping
\be
y_k = \left\{ \begin{array}{lll} t_k y_j  &\mbox{if}&  k\not=j \\ 
y_j  &\mbox{if}& k=j \end{array}\right.
\ee 
Note that to each index there exists a conjugate index defined by the pairing
of Mandelstam variables in the arguments of the K\"allen function, 
$\lambda(y_1y_6,y_2y_5,y_3y_4)$.   
The constraint $\Theta(-\lambda)$ is solved for the variable with index
conjugate to $j$ in each primary sector $j$.
To be explicit we give the result for primary sector 1 in the following, 
where we have
$$\lambda(t_6,t_2t_5,t_3t_4)=0\Leftrightarrow 
t_6^{\pm}=t_2t_5+t_3t_4\pm 2\sqrt{t_2t_3t_4t_5}=
(\sqrt{t_2t_5}\pm\sqrt{t_3t_4})^2
$$

\subsection*{Remapping the square-root singularity}

Before iterated sector decomposition can be applied  
to disentangle the IR singularities, 
it is necessary to perform some variable transformations such that 
finally all possible singularities are at zero. 
Further it is useful to make a quadratic
transformation $t_j=x_j^2$ for $j=2,3,4,5$ to avoid square roots. 

One possibility of  remapping 
is to split each primary sector into two subsectors A  and B, with 
$t_6 \in [t_6^-,t_6^0] = A$
and $t_6 \in [t_6^0,t_6^+] = B$, where $t_6^0=t_2t_5+t_3t_4$. 
The following transformations then lead to a 
numerically stable behaviour near the phase-space boundaries:
\bea
t_6^{A} &=& t_6^- + (t_6^0-t_6^-) x_6^2  = 
( x_2x_5-x_3x_4 )^2 + 2 x_2x_3x_4x_5\, x_6^2\nonumber\\
t_6^{B} &=& t_6^+ - (t_6^+-t_6^0) x_6^2 = 
( x_2x_5+x_3x_4 )^2 - 2 x_2x_3x_4x_5\, x_6^2
\eea 
After these transformations the integral $J_{4,{\rm sec1}}$ in 
primary sector one is given by
\bea
J_{4,{\rm sec1}}&=& \frac{4}{\pi}\,(J_{4,1}^{A}+J_{4,1}^{B})\nonumber \\ 
J_{4,1}^{A,B} &=& 2^{5-2\epsilon} \int\limits_0^1 \prod_{j=2}^6dx_j \,
(x_2x_3x_4x_5)^{1-2\epsilon}\, 
x_6^{-2\epsilon} (2-x_6^2)^{-1/2-\epsilon}
\Theta(1-t_6^{A,B}(\vec x)) \nonumber \\ 
&& \qquad \quad \times F(1,x_2^2,x_3^2,x_4^2,x_5^2,t_6^{A,B}(\vec x))
\eea
The function $F$ depends on the topology, in our case 
\bea
F(z_1,\ldots,z_6) &=& \Bigl( \sum_{j=1}^6 z_j \Bigr)^{-3+4\epsilon} \,
\frac{(z_1+z_5)(z_2+z_6) -z_3z_4}{z_2 (z_2+z_4+z_6)^2}\;.
\eea
Note that the $\Theta$-function constraint is trivially fulfilled 
whenever one of the variables $x_j$, $j\in \{2,3,4,5\}$ 
goes to zero. 
However, at this point we have not mapped all possible singularities to zero
yet. For example, $t_6^A$ can vanish if $\{x_2,x_3,x_4,x_5\}\to 1, x_6\to 0$. 
In this case another transformation $x_j\to 1-x_j$ is 
made\footnote{It can occur that $x_{j}\to 0$ is singular as well for some 
$j\in \{2,3,4,5\}$. In this case the integration 
region for $x_j$ is split at 1/2, and only after this splitting 
the variables are remapped such that all possible singularities are located at
zero.}. 
 Note that all these transformations are done automatically in our program.
Having finally mapped all possible singularities to zero, iterated sector 
decomposition can be applied straightforwardly. One obtains the 
coefficients of the $1/\epsilon$ poles as finite parameter integrals which 
can be integrated numerically, in the
same way as has been explained in\,\cite{Gehrmann-DeRidder:2003bm,Binoth:2000ps}. 

\subsection*{Numerical solution}

For the integral $J_4$ we find numerically
\bea
(1-\epsilon)^2 J_4 = 
 0.500 \frac{1}{\epsilon^2} + 2.749 \frac{1}{\epsilon} + 7.869 + {\cal O}(\epsilon)  
\eea
Inserting this result into (\ref{EqC4}) and 
comparing to eq.\,(\ref{c23}) we see that the poles are cancelled 
within the numerical precision which we chose to be 0.1\%.

\section{Another Topology}

To show that our method also works in the case of 
a more complicated pole structure and a lengthier numerator, 
we consider now the cut graph shown in Fig.\,\ref{fig:topo2}, 
which leads to poles up to $1/\epsilon^4$, and the expanded numerator 
contains about 70 terms. 
\unitlength=1mm
\begin{figure}
\begin{picture}(120,30)
\put(45,0){\epsfig{file=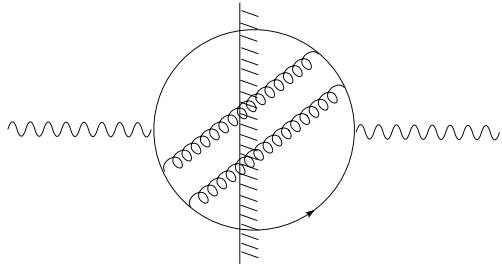,height=3.5cm}}
\end{picture}
\caption{Cut topology with $1/\epsilon^4$ poles.\label{fig:topo2}}
\end{figure}
The corresponding matrix element is given by ($p_{ij\ldots}=\not p_i+\not
p_j+\ldots$)
\bea\label{mtopo2}
| {\cal  M}_{1\to 4} |^2&=& -(4\pi)^3  \alpha \,\alpha_{s}^2 \,C_F^2
\left(\frac{\mu^{2\epsilon}}{\Gamma(1+\epsilon)(4\pi)^\epsilon}\right)^3 
\frac{{\rm tr}(p_1\gamma^\sigma p_{13} \gamma^\rho p_{134} \gamma^\mu p_2 \gamma_\rho
p_{24}\gamma_\sigma p_{234} \gamma_\mu  )}{s_{134}s_{13}s_{234}s_{24}} \nonumber\\
\eea
Writing again  the phase space integral  in the  form
\bea
\tilde{T}_{1\to 4} &=& \int d\Phi_{1\to4} | {\cal M}_{1\to 4} |^2= 
( z_1\, \epsilon )^2\, T_{1\to 2}\,\left( \frac{Q^2}{\mu^2} \right)^{-2\epsilon}
 \frac{1}{16(1-\epsilon)\Gamma(1+\epsilon)^2\Gamma(1-2\epsilon)} \, 
  \tilde J_4(\epsilon)
\nonumber
\eea
we obtain numerically
\bea
\tilde J_4 &=&  \frac{1.000}{\epsilon^4}
           + \frac{3.000}{\epsilon^3}
           + \frac{3.776}{\epsilon^2}
           + \frac{8.957}{\epsilon}
           + 57.85 + {\cal O}(\epsilon)  
\eea
We stress that no analytical manipulations whatsoever, 
in particular no reduction to master integrals, 
are necessary to achieve this result. 

\section{Discussion and Outlook}
In this paper we have argued that the method of sector decomposition 
as proposed 
in \cite{Heinrich:2002rc,Gehrmann-DeRidder:2003bm} 
and applied in a similar form in \cite{Anastasiou:2003gr}
can serve to calculate the double real emission part needed for 
NNLO corrections to jet cross sections in massless QCD. 
We have shown explicitly how the IR poles cancel by considering 
one sample topology contributing to $e^+e^-\to 2$\,jets at NNLO.
We calculated all cuts contributing to a 2 jet final state: 
The double real emission part, corresponding to a 4--particle cut,  
is calculated numerically 
by sector decomposition, and all the poles are shown to cancel   
with contributions from the 2--and 3--particle cuts within 
the chosen numerical precision\footnote{Note that the 
method of sector decomposition could also serve to calculate the
2--and 3--particle cut contributions.
In this case the algorithm  for multi-loop integrals as described
in \cite{Binoth:2000ps,Binoth:2003ak} can be applied.}. 
We also calculated numerically the 4--particle cut of a topology 
which has the maximal possible number of IR poles.

The method has several appealing features:
\begin{itemize}
\item The (overlapping) soft/collinear poles are extracted without 
the need to establish a subtraction scheme and to integrate 
analytically over complicated subtraction terms.
\item The finite parts are available as regular functions of the kinematic
invariants. If a measurement function is included, as already has been 
done in \cite{Anastasiou:2003gr}, one can obtain fully differential 
cross sections, such that for example the implementation of experimental cuts 
should be straightforward.
 
\item It is justified to expect that the Monte Carlo integration to 
obtain the final cross section does not lead to major instabilities 
as the singularities at phase space boundaries already have been 
remapped. 
\item The generalisation of the method to the $1\to 5$ phase space is feasible 
but has to be investigated further.
\end{itemize}
The complete double real emission part contributing 
to $e^+e^-\to 2$\,jets can be evaluated 
along the same lines and will be given elsewhere.

\subsection*{Acknowledgement}
We would like to thank the Kavli Institute for Theoretical Physics
for its kind hospitality while part of this work has been completed.
This research was supported in part by the National Science Foundation 
under Grant No.\,PHY99-07949.


\section*{Appendix}
\subsection*{The D-dimensional $1\to N$ phase space for $N=2,3,4$}

The differential form of the phase space for a  $1\to N$ particle 
phase space in $D$ dimensions is given by
\be\label{Eq:caseN}
d\Phi_{1\to N} = (2\pi)^{ N - D (N-1)} 
\Bigl[ \prod\limits_{j=1}^{N} d^Dp_j \delta(p_j^2) \Theta(E_j)\Bigr] 
\delta\Bigl(Q-\sum\limits_{j=1}^{N} p_j \Bigr)
\ee
Here $Q$ is the incoming momentum and the $p_j$ denote outgoing
particles with light-like momenta and energy component $E_j$. 
As $Q$ is time-like for physical kinematics
one can always achieve $Q=(E,\vec 0^{(D-1)})$ by an adequate Lorentz boost.
Let us specialise now to the cases $N=2,3,4$.

\subsection*{Case $1\to 2$:}
For $N=2$ the momenta can be parametrized by 
\bea
Q   &=& (E,\vec 0^{(D-1)})\;,\;        
p_1 = E_1\, (1,\vec 0^{(D-2)},1) \;,\;  
p_2 = Q-p_1
\eea
Integrating out the $\delta$-distributions leads to
\bea
d\Phi_{1\to 2} = (2\pi)^{ 2 - D} \;2^{1-D}\;(Q^2)^{D/2-2}\; d\Omega_{D-2} 
\eea
where $d\Omega_{D-2}$ is the differential surface element of the 
$S_{D-2}$ sphere.
Its integral is equal to the volume of the $(D-1)$-dimensional unit sphere
\be
\int\limits_{S_{D-2}} d\Omega_{D-2} = 
 V(D-1) = \frac{2\,\pi^{\frac{D-1}{2}}}{\Gamma(\frac{D-1}{2})}
\ee

\subsection*{Case $1\to 3$:}
For $N=3$ one can choose a coordinate frame such that 
\bea
Q   &=& (E,\vec 0^{(D-1)})        \nonumber\\
p_1 &=& E_1\, (1,\vec 0^{(D-2)},1) \nonumber\\
p_2 &=& E_2\, (1,\vec 0^{(D-3)},\sin\theta,\cos\theta)\nonumber\\
p_3 &=& Q-p_2-p_1
\eea
Integrating out the $\delta$-distributions yields
\bea
d\Phi_{1\to 3} =\frac{1}{4} (2\pi)^{3- 2\,D} \; dE_1 dE_2 
d\theta_1 [E_1 E_2\sin\theta]^{D-3}
d\Omega_{D-2} \; d\Omega_{D-3}
\eea
As in the following a parametrization in terms of the Mandelstam variables
$s_{ij}=2\,p_i\cdot p_j$
will be useful, we make the transformation 
$E_1, E_2, \theta \to s_{12},s_{23},s_{13}$.
To work with dimensionless variables we define $y_1=s_{12}/Q^2$, $y_2=s_{13}/Q^2$, $y_3=s_{23}/Q^2$
which leads to
\bea
d\Phi_{1\to 3} &=& (2\pi)^{3- 2\,D}\, \frac{2^{4-D}}{32} \; (Q^2)^{D-3}\;d\Omega_{D-2} \; d\Omega_{D-3}\,[y_1\,y_2\,y_3]^{D/2-2}\nonumber\\&&
 dy_{1}\, dy_{2}\, dy_{3}\, \Theta(y_1)\, \Theta(y_2)\, \Theta(y_3) \,\delta(1-y_1-y_2-y_3)
\eea

\subsection*{Case $1\to 4$:}
Starting from Eq. (\ref{Eq:caseN}) and eliminating $p_4$ yields
\bea
d\Phi_{1\to 4} &=& (2\pi)^{4- 3D}\,\frac{d^{D-1}p_1}{2\,E_1}\frac{d^{D-1}p_2}{2\,E_2}\frac{d^{D-1}p_3}{2\,E_3}\,
\Theta(E_1)\,\Theta(E_2)\,\Theta(E_3)\nonumber\\&&
\Theta(E-E_1-E_2-E_3)\,\delta\Bigl((Q-p_1-p_2-p_3)^2\Bigr)
\eea
Choosing a frame where
\bea
Q   &=& (E,\vec 0^{(D-1)})        \nonumber\\
p_1 &=& E_1\, (1,\vec 0^{(D-2)},1) \nonumber\\
p_2 &=& E_2\, (1,\vec 0^{(D-3)},\sin\theta_1,\cos\theta_1)\nonumber\\
p_3 &=& E_3\, (1,\vec 0^{(D-4)},\sin\theta_3\sin\theta_2,\cos\theta_3\sin\theta_2,\cos\theta_2)\nonumber\\
p_4 &=& Q-p_1-p_2-p_3\;\;.
\eea
leads to
\bea
d\Phi_{1\to 4} &=& \frac{1}{8}(2\pi)^{4- 3\,D}\,dE_1\,dE_2\,dE_3\,d\theta_1\,d\theta_2\,d\theta_3
[E_1E_2E_3\sin\theta_1\sin\theta_2]^{D-3}\sin\theta_3^{D-4}\nonumber\\
&&\, d\Omega_{D-2} \; d\Omega_{D-3}\; d\Omega_{D-4}
\Theta(E_1)\,\Theta(E_2)\,\Theta(E_3)\Theta(E-E_1-E_2-E_3)\nonumber\\
&&\delta(E^2-2E(E_1+E_2+E_3)+2(p_1\cdot p_2+p_1\cdot p_3+p_2\cdot p_3))\label{a11}
\eea
As above we map the angle and energy variables to 
the Mandelstam invariants as integration variables.
The Jacobian in 
combination with terms already present in (\ref{a11}) 
can be written as the determinant of the Gram matrix $G_{ij}=2\,p_i\cdot p_j$.
The determinant can be expressed by the K\"allen function 
$\lambda(x,y,z)=x^2+y^2+z^2-2xy-2yz-2xz$ as
\bea
\det(G) &=& \lambda(  s_{12}\,s_{34}, s_{13}\,s_{24},s_{14}\,s_{23} )\nonumber\\
        &=& -[ 4\,E\,E_1\,E_2\,E_3\,\sin\theta_1\,\sin\theta_2\,\sin\theta_3 ]^2
\eea
We see that $\det(G)$ has to be negative semi-definite.
With the dimensionless variables
$$
y_1 = s_{12}/Q^2\,,\, y_2 = s_{13}/Q^2\,,\,y_3 = s_{23}/Q^2\,,\,y_4 = s_{14}/Q^2\,,\,y_5 = s_{24}/Q^2\,,\,y_6 = s_{34}/Q^2
$$
and $\lambda = \lambda(y_1y_6,y_2y_5,y_3y_4)$ one obtains finally 
\bea
d\Phi_{1\to 4} &=& (2\pi)^{4- 3\,D} (Q^2)^{3 D/2-4}\, 2^{-2D+1}\, d\Omega_{D-2} \; d\Omega_{D-3}\; d\Omega_{D-4}\,\nonumber\\&&
\left[ \prod\limits_{j=1}^{6} dy_j \Theta(y_j) \right] \,
\Theta(-\lambda)\,[-\lambda]^{(D-5)/2}
\delta(1-\sum_{j=1}^{6} y_j)\label{fi4}
\eea

\end{document}